# Infrastructure-enabled GPS Spoofing Detection and Correction

Feilong Wang[†], Yuan Hong[‡], Jeff Ban[†,*]

[†]University of Washington, WA, United States

[‡]University of Connecticut, CT, United States

[*]Corresponding author: banx@uw.edu

**Abstract**

Accurate and robust localization is crucial for supporting high-level driving automation and safety. Modern localization solutions rely on various sensors, among which GPS has been and will continue to be essential. However, GPS can be vulnerable to malicious attacks and GPS spoofing has been identified as a high threat. With transportation infrastructure becoming increasingly important in supporting emerging vehicle technologies and systems, this study explores the potential of applying infrastructure data for defending against GPS spoofing. We propose an infrastructure-enabled framework using roadside units as an independent, secured data source. A real-time detector, based on the Isolation Forest, is constructed to detect GPS spoofing. Once spoofing is detected, GPS measurements are isolated, and the potentially compromised location estimator is corrected using secure infrastructure data. We test the proposed method using both simulation and real-world data and show its effectiveness in defending against various GPS spoofing attacks, including stealthy attacks that are proposed to fail the production-grade autonomous driving systems.

## I. INTRODUCTION

Technologies supporting advanced driving systems have been evolving at an unprecedented pace in recent years. Among them, accurately localizing a vehicle's global positions is critical for its core role in vehicle routing and control. To support high-level driving automation and safety, localization modules must be robust in various driving scenarios, which demand advanced sensors and algorithms. Modern localization modules rely on multiple sensors, including, for example, Global Positioning System (GPS), Inertial Measurement Unit (IMU), Light Detection and Ranging (LiDAR), and camera [1]. However, sensors on vehicles are vulnerable to malicious attacks [2]. For example, GPS spoofing, which broadcasts falsified GPS signals, has been a long-recognized high threat [3]; LiDAR can be compromised by replay attacks that deceive receivers with recorded (thus outdated) data [4]; cameras are sensitive to blinding attacks that emit light into the camera [5]. Despite the disclosed vulnerabilities, solutions to addressing them are still limited.

This study focuses on GPS spoofing detection and correction as GPS has been and will continue to be an essential technique for vehicle localization [6]. Attacks on GPS have long been recognized, notably jamming, replaying, and spoofing attacks. GPS jamming prevents the receivers from receiving signals properly, while a replaying attack records authentic signals and replays the outdated or irrelevant signals to interrupt the proper operation of vehicles. GPS spoofing aims to forge signals to mislead a vehicle to deviate from its planned path [9, 10], hence endangering the safety of passengers and other road users. Section II-A provides a detailed comparison of these attacks. Being a false data injection attack (see a full taxonomy of attacks in [2]), GPS spoofing can be the most effective among the three types of attacks as it allows the attacker to dictate the victim vehicle's positions to achieve specific goals [9]. Despite being a real threat, defending against GPS spoofing is still an open security problem from both *prevention* and *detection* perspectives [3]. From the prevention perspective, a fundamental measure to prevent GPS spoofing is to apply cryptographic techniques to civilian GPS infrastructure [10]. However, it requires considerable modifications or even reconstruction of the existing satellite infrastructure and GPS receivers, which is impractical. From the detection perspective, the defense methods vary by the source of information used for detecting malicious attacks. The classical techniques are based on collecting and analyzing GPS signals in real time, such as accurate clock information or angle of arrival [9], [11]. Though effective, these techniques may not be generalizable as each technique is designed for specific attacks and may need a large budget for installing dedicated devices (e.g., multiple antennae) on individual vehicles. Another open question is how to recover accurate navigation after an attack is detected [10].

With various sensors increasingly prevalent in vehicles, detecting sensor (e.g., GPS) attacks via cross-comparing multiple data sources has attracted considerable attention in recent years [12]–[14]. One typical approach is to detect anomalies in received real-time measurements by comparing them with patterns in previously recorded data. This is often done by a supervised machine learning model or a statistical model corresponding to specific attacks from these records and applying the learned model to real-time anomaly detection [15]. One challenge of implementing such methods is the lack of labeled records for model training or the imbalance between benign and adversarial records. To address this challenge, some studies have developed methods based on one-class classification where anomaly detection models are trained using only benign data [16]. In practice, the methods may be difficult to implement as extracting features for one-class classification training is not trivial [16]. Another type of approaches to detect attacks is to integrate other real-time data sources (e.g., IMU data) with the vehicle's (mathematical) motion model [6]. An anomaly/attack is detected if data from the subject sensor deviate too much from the predicted output (e.g., vehicle's position) from the motion model [17]. However, the motion model could be compromised when GPS spoofing occurs (and before spoofing is detected), leading to unreliable predictions [18]; see also the numerical results and discussions later in Section VI-B. One mitigation is to simultaneously run multiple models on redundant sensors (e.g., GPS, LiDAR and camera) and detect attacks via cross-validation. Yet, implementing and cross-validating multiple models can be complicated, especially for identifying the attack source [14, 19]. Installing multiple redundant sensors can also be costly, given the vast number of vehicles on roads and constrained onboard resources.

Besides emerging vehicle-based sensors and technologies, transportation infrastructure is becoming increasingly important in supporting various functionalities of advanced vehicle technologies, especially Connected and Automated Vehicles (CAVs) [19]–[21]. It is widely accepted now that infrastructure-vehicle cooperation is probably a more viable path to implement emerging systems, e.g., automated driving, compared with that using driverless vehicle technologies solely. For this, the communication and data transmission between vehicles and infrastructure will play a central role. Indeed, V2X messages (e.g., the basic safety message (BSM)) have already been defined for data transmitted between vehicles and "everything" (including other vehicles, the infrastructure, and other users of the roadway), and secure data transmission schemes (e.g., the secure credential management system (SCMS) [22]) have also been proposed for V2X data. Emerging V2X communication systems, such as 5G-based Cellular V2X, are capable of supporting real-time decisions in, for example, collision avoidance systems and positioning of vehicles. Leveraging secure data from the infrastructure may help defend against cybersecurity attacks, including GPS spoofing attacks. Therefore, while we should continue to encourage research on more effective GPS spoofing defense methods based on signal processing, anomaly detection, and data fusion (some recent methods can be found in [8], [13], [23]), we should also welcome methods via exploring the use of secure infrastructure data for GPS spoofing detection and mitigation.

This study focuses on such a new exploration by proposing an infrastructure-enabled defense (IED) framework via utilizing roadside units (RSU) as an independent, secure data source. An RSU broadcasts locational information (similar to or could be part of the V2X data from RSU); vehicles in the broadcast range can use the information to estimate their locations periodically (see Section V-A for more details). Such secure, independent data from RSUs enables new ways to detect and mitigate GPS spoofing, which we will explore and elaborate more in the remainder of this paper. The proposed IED framework has several unique features compared with existing solutions. First, it takes advantage of the communication modules between vehicles and infrastructure (e.g., existing or newly deployed V2X devices), instead of requiring sophisticated in-vehicle GPS receivers or redundant sensors for cross-validation. Second, enabled by the secure data from infrastructure, it is feasible to design a simpler yet effective defense solution to detect and correct GPS spoofing. Computed from secure RSU data, the features for attack detection are also "protected" (i.e., safe from attackers' manipulation), relieving the challenge of developing attack-resilient algorithms [24]. Third, it is more practical to secure the information from RSUs than to secure the established civilian GPS satellite infrastructure (see Section II-D for more discussions). Therefore, the proposed IED solution provides a new and valuable alternative to addressing GPS spoofing issues. Furthermore, exploring IED solutions for GPS spoofing may provide helpful insights to address other data-related cybersecurity issues in transportation, which we will elaborate more in later sections. We note here that, while we focus on GPS spoofing on ground vehicles in this paper, GPS spoofing has also been studied for aircraft and marine vehicles (ships) [7]. In fact, an infrastructure-based GPS spoofing mitigation idea for aircraft was also reported in [25]. However, due to the distinct characteristics/operations of ground vehicles and aircraft (or ships), their safety requirements, and the drastically different space they are operated in, methods for aircraft or ships cannot be applied directly to ground vehicles (e.g., the idea in [25] does not apply to ground transportation).

We first introduce the design of secure RSU data and the method of how a vehicle interacts with the infrastructure to obtain secure, global position measurements. Based on the secure measurements, we develop and compute multiple features, with which a real-time detector, based on the Isolation Forest, is constructed to detect GPS spoofing. Once spoofing is detected, GPS measurements are isolated, and the potentially compromised location estimator is corrected using the RSU data. We design the detection and correction methods under the situation that RSU data is not always available due to certain constraints (e.g., a limited budget to install RSUs all over the road network). If RSU data are not available, an RSU-based prediction model utilizes the last available RSU measurement and the vehicle motion model to predict vehicle locations, preserving timely attack detection. We test the proposed IED framework using both simulation and real-world data and show its performance compared with state-of-the-art solutions in defending various types of GPS spoofing, including a stealthy attack that is proposed to fail the production-grade autonomous driving systems [16]. The major contributions of this paper are summarized as follows.

1) This study explores and proposes an IED framework for detecting and correcting GPS spoofing that complements existing methods that mainly rely on (likely insecure) vehicular data.
2) By integrating both GPS and RSU data, we develop a machine learning-based spoofing detection method that is simple yet effective in detecting GPS spoofing, adding new tools to the current toolbox for GPS spoofing.
3) A new correction model is also developed leveraging the RSU data, which results in much-reduced location errors when GPS spoofing attacks occur.

In the rest of this paper, we review related works in Section II. In Section III, we present the problem statement and major assumptions and some preliminaries on which our problem is constructed. Section IV introduces GPS spoofing attack models. Section V presents the proposed IED framework and Section VI evaluates it using both simulation and real-world data. Concluding remarks are discussed in Section VII.

## II. LITERATURE REVIEW

### *A. GPS Spoofing Attacks*

Existing studies have revealed potential vulnerabilities of localization sensors to malicious attacks [13], [16]. GPS is particularly prone to attacks, including jamming, replaying, and spoofing [2]. GPS jamming can prevent vehicles from receiving GPS signals properly by, e.g., transmitting radio signals that overpower the (weak) authentic GPS signals. Jamming could be addressed by implementing beam/null-steering antenna arrays that can filter out jamming signals [26]. Replaying attacks aim to confuse vehicles by recording and rebroadcasting GPS signals that could be outdated or irrelevant to the vehicles' real-time operation. False signals in such attacks could be identified by monitoring the receiver's clock bias over time [26]. GPS spoofing misleads vehicles' trajectories by forging counterfeit GPS signals, which could be done by intercepting and falsifying authentic signals before sending them to GPS receivers [3]. GPS spoofing falls into the broad category of false data injection attacks, which compromise sensor readings stealthily so that undetected errors are introduced into state predictions. A full taxonomy of various types of attacks can be found in [2]. It is well recognized that GPS spoofing can be stealthy to be detected among these attacks and is still an open challenge in the cybersecurity community.

Before discussing existing defense solutions against GPS spoofing, we summarize common types of GPS spoofing in recent studies[10], [13], [14].

- *Instant*: One GPS measurement that is unexplainable and significantly different from previous ones.
- *Noise*: A consecutive sequence of GPS measurements with increased variance. Noise attack occurs across multiple successive sensor readings.
- *Constant bias*: A sequence of GPS measurements with a constant offset from the vehicle's true locations.
- *Gradual drift (stealthy attack)*: A sequence of GPS measurements that are modified to gradually deviate the vehicle from its true trajectory during a period of time.

The references above also discuss in detail the consequences of each type of GPS spoofing attacks. Among these attacks, the constant bias and gradual drift attacks have received the most attention. In particular, the gradual drift attack is one type of stealthy attacks, which is more deceptive than other attacks: it can result in a large deviation between the true trajectory and the falsified trajectory over time. Sophisticated stealthy attacks have been proposed in recent studies, making them difficult to be detected. For example, stealthy GPS spoofing is proposed in [13] to gradually drift the true vehicle position according to its kinematic model. In [1], a stealthy GPS spoofing attack (named FusionRipper) is designed to fail production-grade autonomous driving systems (e.g., Baidu's Apollo system) with an over 90% success rate. FusionRipper targets the predominantly adopted Multi-Sensor Fusion (MSF) algorithms and performs exponential spoofing, which injects mild deviations at the beginning to gradually compromise MSF and then aggressive deviations with exponential growths. The deviations injected over time are controlled by two parameters which are tuned according to MSF's configuration. In this study, we implement FusionRipper as a stealthy attack to test the IED framework.

### *B. Detection Methods against GPS Spoofing*

Defending GPS spoofing could be done from the prevention perspective, i.e., enhancing data security via techniques such as encryption and user authentications. Preventing GPS spoofing this way requires significant modifications of the civilian GPS satellite infrastructure (i.e., satellites, GPS receivers, and their communication that is currently without any encryption scheme) that has been widely deployed and used for decades. Clearly, doing so would be very costly and impractical [3]. As well recognized and adopted extensively in previous studies [9], [11], [27], practical GPS spoofing defense solutions contain two major steps: spoofing *detection* and spoofing *correction* (mitigation). We review detection methods here, while correction methods are covered in the next subsection.

Classical GPS spoofing detection methods focus on collecting and processing rich information in GPS signals, such as accurate clock information, signal power and arrival angle [1], [9], [23]. These methods have been shown effective in detecting specific types of attacks. However, they often require dedicatedly designed GPS receivers in vehicles (e.g., receivers with moving or multiple antennae) and may not be generalizable to sophisticated attacks that largely mimic authentic GPS signals [3]. Meanwhile, how to correct the compromised location estimator and recover accurate localization after attack detection is still an open question [10].

In recent years, sensors are increasingly installed in vehicles and this has promoted studies that detect spoofing attacks (i.e., anomalies) via cross-validating multiple data sources [12], [13], [16]. Such studies can be categorized into two groups: *data-driven* and *model-based* [16]. The former relies on prepared (historical) data to learn a set of patterns or rules, with which the real-time sensor data is determined as benign or adversarial [12], [14]. The rules could be learned by formulating a supervised learning problem, where a classifier is learned using the labeled training data. The trained classifier serves as the detector to detect whether a sensor is under attack or not [15]. Such supervised learning algorithms have been shown effective in detecting spoofing attacks on real-time localization systems implemented on a wheeled robot [28]. Recently, deep learning methods have been applied to detecting anomalies in speed sensors [14]. Despite their success in specific applications, supervised-learning-based methods have two limitations [15]: i) the training data requires labeled (at least two classes of) records, which can be challenging to prepare; and ii) the trained model may not be generalizable to address new types of attacks that are not represented in the training data. To address these limitations, recent studies propose to perform unsupervised learning or one-class classifications (OCC) that are trained only on normal data and thus do not require specific labels associated with the data [29]. Then, real-time sensor data is fed into the learned classifier to detect attacks or anomalies. In [7], the authors formulated attack detection as an unsupervised binary classification problem and applied K-means to cluster the data into two groups, one for attack and the other for non-attack. Applying K-means to detect stealthy GPS spoofing can be challenging, as it requires predetermining the feature space and distance function for measuring the distance between data points. In [13], a One-Class Support Vector Machine (OCSVM) model is proposed to detect anomalies in vehicular sensor readings. Though robust in detecting inconsistencies among data sources, studies have shown that OCSVM could be sensitive to outliers and tends to produce false-positive errors [30]. Meanwhile, OCC-based detectors do not address another limitation associated with the data-driven methods: the detector may detect the existence of anomalies but could fail to identify their source (i.e., which sensor is under attack). This may make it challenging to design and implement mitigation measures (e.g., isolating the attacked sensor).

Model-based detection methods involve modeling and continuously predicting a vehicle's motion dynamics using real-time measurements from the vehicle [31], [32]. The basic idea is that if a sensor measurement deviates from the expected value from the vehicle dynamic model too much, the sensor may be compromised. The $\chi^2$-test-based detection is often used to determine whether the deviation is large enough to claim the sensor being an outlier or under attack [33]. The detection test is a statistical test, based on the statistic Normalized Estimation Error Squared (*NEES*) that follows a $\chi^2$ distribution [8], [34]. The $\chi^2$-test-based detection can be sensitive to sensor noises, resulting in a high rate of false positives (i.e., outliers that are incorrectly identified as attacks due to sensor noises). To mitigate this issue, a cumulative sum (CUSUM) discriminator is recently proposed to detect attacks on GPS and LiDAR [16]. CUSUM detects an attack by inspecting multiple consecutive sensor measurements instead of one measurement only: if the inconsistency between the sensor measurement and the expected vehicle position appears continuously, the sensor is likely under attack. There are some limitations with CUSUM in real-world applications. First, it requires two tuning parameters that can be challenging to determine in real-world implementations. Second, being a model-based method, it relies on a prediction model that may be compromised by stealthy attacks. Specifically, an attack can carefully manipulate the input to the prediction model such that the generated predictions are corrupted. If this occurs, the features computed from the predictions are no longer reliable indicators of attacks. In the numerical experiments in this paper, we show the weakness of CUSUM when facing stealthy attacks.

### *C. Mitigation/Correction Methods again GPS spoofing*

Existing studies are mainly on attack detection and have limited discussions on mitigating/correcting the errors caused by the attack [28], [31]. The typical strategy is to run a fail-safe mechanism (e.g., handing over control to the human driver) if an attack is detected [35]. However, such a fail-safe mechanism can be costly as it interrupts the system or may not be applicable in certain scenarios (e.g., automated driving).

Another typical solution is to deploy multiple sensors, such that an attacked sensor is isolated and the system relies on the rest of the sensors [36]. For example, a vehicle equipped with GPS and LiDAR will rely on LiDAR for localization if GPS spoofing is detected [16]. However, there are some limitations to such solutions. First, as noted above, identifying the attack source (i.e., which sensor is under attack) in the multi-sensor setting is often challenging, especially when all sensors are vulnerable. Consequently, isolating the attacked sensor is not trivial. Second, in the presence of detection lag, the data fusion framework would have been partially compromised before noticing an attack and isolating the attacked sensor [1]. Previous studies only emphasize isolating the attacked sensor but lack discussions on correcting the compromised data fusion framework. One possible solution is to run a secondary system (e.g., a localization module independent of GPS sensor) so that the system under attack is isolated and replaced by the secondary system [35]. Yet, deploying and running redundant systems could be economically and computationally costly.

### *D. Methods of Obtaining Secure Infrastructure Data*

Infrastructure plays an increasingly important role in modern driving systems, facilitating their various advanced functions, such as detecting pedestrians and efficient driving at intersections [12], [37]. The proposed IED framework in this paper requires secure infrastructure data (RSU data). Yet, the infrastructure data itself can be vulnerable to malicious attacks, including DoS attacks and spoofing attacks. Fortunately, active research has been conducted on securing infrastructure and practical security strategies are currently available [38].

Infrastructure data collection and transmission can be secured by applying a variety of state-of-the-art secure channels that use advanced encryption algorithms (e.g., DES, 3DES, AES, RSA and Blowfish [39]). These existing encryption methods can be evaluated in transportation applications and revised, if needed, to fit transportation scenarios better. In practice, secure data communication is becoming a standard in CAV development and deployment. For example, a recent review in [38] summarizes the integrity of V2X communication from different contexts, such as reputation analysis and message integrity checking. In [22], SCMS is presented to secure V2X data. SCMS issues digital certificates to vehicles and RSUs to secure their communications while maintaining efficient revocation of misbehaving or malfunctioning vehicles. SCMS may be readily used for secure data transmission in our proposed IED framework. Besides data transmission, the received secure infrastructure data may also be encrypted before storage (and decoded before using them), ensuring data security even if the system (hardware) is hacked [40].

These existing studies suggest that secure data transmission between vehicles and the infrastructure can be reasonably done. As we focus on developing GPS spoofing detection and mitigation methods using infrastructure data, in this paper, we apply the state-of-the-art encryption method to set up secure channels for secure data transmission between an RSU and its nearby vehicles. Specifically, we implement an Advanced Encryption Standard (AES) scheme [41] in terms of the process of encrypting and decrypting transmitted data with user authentication, which is similar to the SCMS scheme for secure V2X transmission. See Section V-A and Section VI-A for more detailed discussions on this.

## III. PROBLEM STATEMENT AND PRELIMINARIES

### *A. Problem Statement*

Fig. 1 illustrates the problem setup and the general idea of the IED framework against GPS spoofing. We consider a simple yet common localization solution, where a vehicle can be tracked by a typical motion model with high-frequency local measurements from a low-end IMU and takes low-frequency global measurements from GPS for correcting location errors periodically. Low-end IMUs are pervasive nowadays and are widely deployed in smartphones and vehicles. The problem setting here ensures the generality of the study since one can obtain IMU measurements from a vehicle's OBD portal [42], without installing additional sensors or utilizing the data from such sensors even if they are installed. GPS could be spoofed in an adversarial environment. The vehicle could deviate from the desired trajectory if spoofing is not detected. Our goal here is to propose an IDE method with which the vehicle can utilize the secure data from RSUs to timely detect GPS spoofing and correct location errors incurred by the attacks. Section V-A provides more details about the data provided by the RSU.

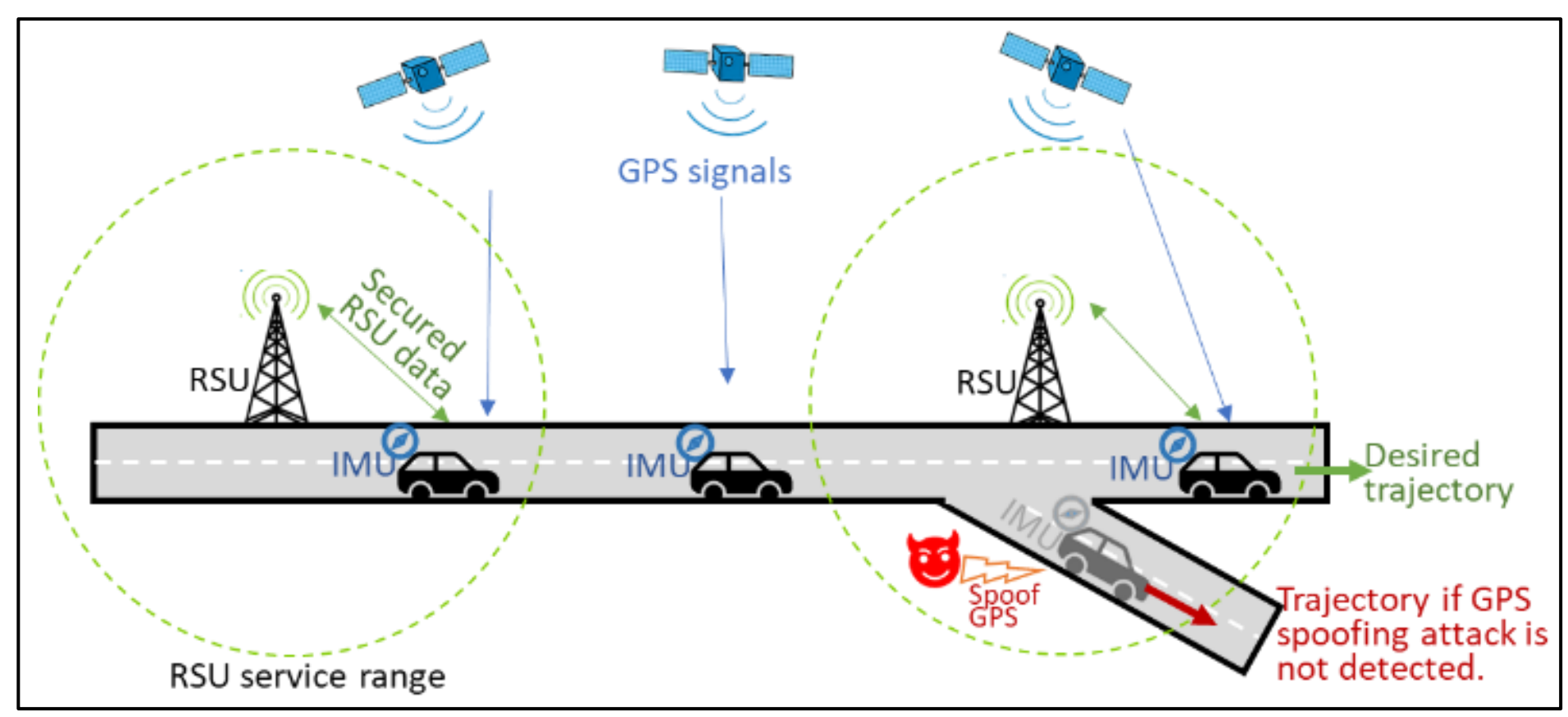

**Fig. 1.** ILLUSTRATION OF GPS SPOOFING AND IED SOLUTION.

*B. Assumptions*

We impose the following assumptions to simplify our discussion and clarify the focus of this study.

1) GPS spoofing studied here belongs to data security, which is orthogonal to attacks/defenses of hacking into software or hardware systems, or physical network security [38], [43]. To focus on the research challenges and methods of GPS spoofing, we assume in this paper that other attacks have been mitigated with proper countermeasures. The only exception is the methods for secure data transmission between vehicles and the infrastructure; see 2) below.
2) Vehicles can obtain secure RSU data to calculate their global locations. As discussed in Section II-D and more in Section V-A, we assume that secure RSU data can be readily available by applying (or tailoring) existing security schemes [38]. This paper directly applies AES [41] to secure the data and focuses on developing and testing detection and correction methods.
3) We assume that IMU is secure due to assumption 1) above. IMU measurements are typically accessed via a wired channel; thus, their exposure to potentially adversarial environments is low unless in the presence of physical attacks against in-vehicle hardware. This assumption has also been widely adopted in recent cybersecurity research involving IMUs [9], [17].

*C. EKF-based Localization Model*

Estimating vehicle positions from multiple sensors can be achieved by a Kalman Filter (KF)-based method or its variants [44]. Here we briefly describe the KF-based localization model used in this paper to combine GPS (global) and IMU (local) data. Vehicle (global) location at time $k$ is represented by the KF's state $\hat{\boldsymbol{x}}_k$ and uncertainty with a covariance matrix $\hat{\boldsymbol{P}}_k$. Due to the non-linearity of the vehicle motion model, we adopt an Extended Kalman Filter (EKF) proposed in [1].

Following initialization at $k = 0$, EKF estimates the vehicle positions by iterating a prediction step and an update step. The prediction step iterates the motion model (1) to predict the vehicle positions using IMU data; the process is often referred to as *dead-reckoning*. This prediction step is expressed as a discretized vehicle motion model (1) together with the propagation of uncertainty (2) [45].

$$\boldsymbol{x}_k = f(\boldsymbol{x}_{k-1}, \boldsymbol{u}_k) \tag{1}$$

$$\boldsymbol{P}_k = \boldsymbol{F}_{k-1}\hat{\boldsymbol{P}}_{k-1}\boldsymbol{F}_{k-1}^T + \boldsymbol{L}_{k-1}\boldsymbol{Q}\boldsymbol{L}_{k-1}^T \tag{2}$$

Here, $\boldsymbol{x}_k$ and $\boldsymbol{P}_k$ represent the vehicle position and its uncertainty at time step $k$, respectively. $\boldsymbol{u}_k$ gives the IMU measurement containing white noises $\boldsymbol{w}_k$ with covariance matrix $\boldsymbol{Q}$. $\boldsymbol{F}_{k-1} = \frac{\partial f_{k-1}}{\partial \boldsymbol{x}_{k-1}}|_{\hat{\boldsymbol{x}}_{k-1}}$, $\boldsymbol{L}_{k-1} = \frac{\partial f_{k-1}}{\partial \boldsymbol{w}_{k-1}}|_{\hat{\boldsymbol{x}}_{k-1}}$ are the partial derivative matrices corresponding to the state and noises that are obtained by linearizing the system model (1).

The update step is for periodically correcting the cumulated errors in the prediction steps once GPS data $\boldsymbol{z}_k^{GPS}$ is received. The measurement model for GPS data is given by [16]:

$$\boldsymbol{z}_k^{GPS} = \boldsymbol{H} \times \boldsymbol{x}_k + \boldsymbol{e}_k^{GPS} \tag{3}$$

Here matrix $\boldsymbol{H}$ maps vehicle position to the measurement space. $\boldsymbol{e}_k^{GPS}$ is the measurement noise which is assumed to be additive white noise with covariance matrix $\boldsymbol{R}^{GPS}$.

As shown in (4), the update step takes a GPS measurement $\boldsymbol{z}_k$ and its uncertainty $\boldsymbol{R}^{GPS}$ as input to compute the Kalman gain $\boldsymbol{K}_k$, which is then used to correct the predicted state [1].

$$
\begin{aligned}
\boldsymbol{K}_k &= \boldsymbol{P}_k \boldsymbol{H}^T (\boldsymbol{H}\boldsymbol{P}_k \boldsymbol{H}^T + \boldsymbol{R}^{GPS})^{-1} \\
\widehat{\boldsymbol{x}}_k &= \boldsymbol{x}_k + \boldsymbol{K}_k \left(\boldsymbol{z}_k^{GPS} - \boldsymbol{H}\boldsymbol{x}_k\right) \boldsymbol{r}_k^{GPS} \\
\widehat{\boldsymbol{P}}_k &= \boldsymbol{P}_k - \boldsymbol{K}_k \boldsymbol{H} \boldsymbol{P}_k
\end{aligned} \tag{4}
$$

## IV. ATTACK MODELS

Attack models are essential for investigating attack detection and mitigation. We consider two types of GPS spoofing attacks: the constant bias attack and the stealthy attack. As shown in the results section, these two attack models allow for evaluating the IED framework under stealthy and non-stealthy attacks, generating some interesting insights. Other types of spoofing attacks on GPS discussed in Section II-A (including instant and noise attacks) are not implemented in this study, since they either fall out of the scope of this study (e.g., DoS attacks) or can be approximated by the constant bias or stealthy attacks [14] (also briefly discussed below).

### *A. Constant Bias Attack*

A constant bias attack injects a constant bias into the true measurements, causing the GPS readings to deviate from the true ones temporarily. In practice, attackers could launch a bias attack to mislead a vehicle by adding a lateral offset or a longitudinal offset (or both) to the true GPS readings $\boldsymbol{z}_k^{GPS}$. Mathematically, the received GPS measurement would be:

$$
\tilde{\boldsymbol{z}}_k^{GPS} = \boldsymbol{z}_k^{GPS} + \boldsymbol{C} \; (k \in [t_s, t_e]), \tag{5}
$$

where $\tilde{\boldsymbol{z}}_k^{GPS}$ is the spoofed GPS data, and $\boldsymbol{C}$ is a constant vector that can be added to the true GPS readings. $t_s$ and $t_e$ represent the start time and end time of the attack, respectively. With a constant bias attack, the vehicle may be deceived by believing that it is at the wrong location on the roadway and thus takes faulty actions. Notice that an instant attack can be implemented by taking $t_e = t_s + 1$.

### *B. Stealthy Attack*

A stealthy attack injects a sequence of increasing deviations into true measurements, such that the vehicle gradually drifts away from its true trajectory. Mathematically, the received GPS measurement can be expressed as:

$$
\tilde{\boldsymbol{z}}_k^{GPS} = \boldsymbol{z}_k^{GPS} + \boldsymbol{c}_k \; (k \in [t_s, t_e]), \tag{6}
$$

where $\boldsymbol{c}_k$ is carefully designed to avoid triggering an attack detector. Stealthy attacks are more deceptive than constant bias attacks; multiple such strategies have been proposed for GPS spoofing. To implement a noise attack, one could generate $\boldsymbol{c}_k$ by sampling a random distribution (e.g., norm distribution) with a large variance.

As noted in Section II-A, we implement FusionRipper, the state-of-the-art stealthy spoofing strategy that is recognized by top-tier cybersecurity communities [1]. In this study, the implementation of FusionRipper is simplified since our localization solution includes no LiDAR as in the original study. Specifically, we skip the vulnerability profiling step (for determining when GPS measurements dominate the location estimator) and implement the aggressive spoofing step directly. The aggressive spoofing performs exponential spoofing that increases the deviation $\boldsymbol{c}_k$ exponentially. As shown by (7), the deviation $\boldsymbol{c}_k$ is a function of time $k$, controlled by two parameters: $m$ and $n$ (with $n$ slightly larger than 1). At the beginning of the attack, the deviation is small, making it difficult to be detected. As a result, the spoofed GPS measurements would be fused and corrupt the data fusion framework (i.e., EKF). Once this occurs, aggressive deviations can be injected without alerting the detection algorithm.

$$
\boldsymbol{c}_k = m * n^k \tag{7}
$$

## V. INFRASTRUCTURE-ENABLED DEFENSE METHOD

An overview of the IED framework is shown in Fig. 2. Besides the EKF-based localization model that continuously localizes the vehicle (Section III-C), there are three new components. The first component aims to obtain secure, global measurements of vehicle positions from RSUs. The second one (RSU-enabled detection component) runs a real-time detector to monitor whether a received GPS measurement is spoofed or not. The third component is to correct the vehicle location using RSU data. In the following, we describe each of the three components in detail.

### *A. Secure RSU Data from the Infrastructure*

#### 1) **Design of Secure RSU data**

Methods for obtaining secure RSU data include two major aspects: (i) what data to collect and how to collect them; and (ii) how to secure data collection and transmission. We focus on (i) in this study. For (ii), as discussed in Section

II-D, we apply the AES scheme, one of state-of-the-art encryption methods, to design dedicated secure channels for secure data collection and transmissions, focusing on testing its performance in spoofing detection and correction in Section VI.

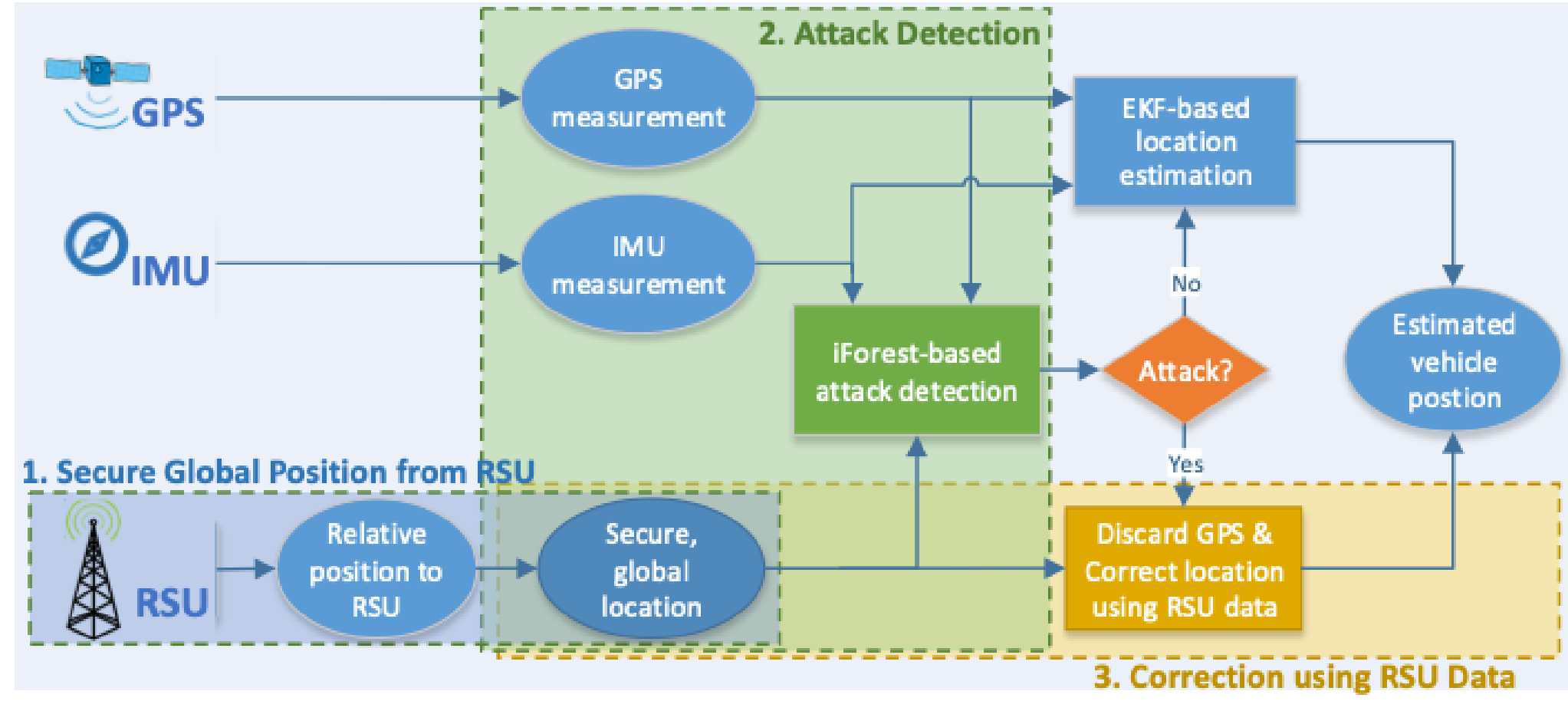

**Fig. 2.** IED SOLUTION FOR GPS SPOOFING DETECTION AND CORRECTION.

The design of secure RSU data ensures that a vehicle can use the data to obtain a global position measurement similar to GPS, denoted as $\boldsymbol{x}_k^{RSU}$. This has been extensively studied in the field of GPS-free localization [46]–[48]. A common practice is to first estimate the vehicle's relative position to the RSU via ranging methods and then compute the vehicle's global position given the (global) coordinates of the RSU [38]. In a ranging method, the distance between a radio transmitter (the RSU here) and a receiver can be inferred from the properties of the radio wave observed at the receiver [47]. Note that this distance is termed as *range* following the literature. The widely known ranging methods include those collecting and utilizing *received signal strength (RSS)*, arrival time or arrival angle [46]. For CAVs that can communicate with RSUs, such range information can be readily available on the vehicle side. Following [46], we use $M(\bullet)$ to express a ranging method that obtains the range information $z_k^{RSU}$ at time $k$:

$$z_k^{RSU} = M(\boldsymbol{x}_k, Crd^{RSU}) + e_k^{RSU}. \tag{8}$$

Here, $M(\bullet)$ is essentially a measurement model depending on the vehicle's (true) global position $\boldsymbol{x}_k$ and the RSU's coordinates $Crd^{RSU}$. $e_k^{RSU}$ is the measurement noise in a Gaussian distribution with covariance matrix $\sigma^{RSU}$. In [46], a recent review of RSU-assisted localization methods is provided, which vary with the RSU data types and configurations of signal transmitters on RSUs and receivers on vehicles. There are also real-world implementations in GPS-absent environments (e.g., Waze's Beacon program to provide navigation for drivers underground [49]). The RSU-assisted localization methods could reach an accuracy in centimeters, much higher than that of GPS [47].

In this study, we implement an efficient and low-cost V2X-based vehicle localization method by Ma et al. [50]. See a discussion of its efficiency in terms of computational latency below. It is low cost as it needs only a single data transmitter on the RSU side and a single receiver on the vehicle (i.e., it is similar to and can be implemented via the current V2X framework), compared with other ranging methods using multiple transmitters or receivers to collect information such as angle of arrivals [50]. Ma et al. [50] assumes that the RSU broadcasts its coordinates, and a vehicle receives the message and extracts associated range information (i.e., the relative distance information). Then the vehicle computes its global position $\boldsymbol{x}_k^{RSU}$ using a sequence of range information $z_k^{RSU}$. Therefore, this method may be readily deployed based on the current V2X systems without additional hardware requirements (the range information does need to be extracted from the receiver on the vehicle side). Omitting the details, we denote this method with function $G(\bullet)$ [50]:

$$\left(\boldsymbol{x}_k^{RSU}, \boldsymbol{R}_k^{RSU}\right) = G\left(\left[\boldsymbol{z}_k^{RSU}, \boldsymbol{z}_{k-1}^{RSU}, \dots, \boldsymbol{z}_{k-o}^{RSU}\right], Crd^{RSU}, \left[\boldsymbol{u}_k, \boldsymbol{u}_{k-1}, \dots, \boldsymbol{u}_{k-o}\right]\right). \tag{9}$$

$\left[\boldsymbol{z}_k^{RSU}, \boldsymbol{z}_{k-1}^{RSU}, \dots, \boldsymbol{z}_{k-o}^{RSU}\right]$ is the sequence of range information associated with the messages from an RSU. $\left[\boldsymbol{u}_k, \boldsymbol{u}_{k-1}, \dots, \boldsymbol{u}_{k-o}\right]$ is a sequence of local measurements containing either speeds or local displacements. These local measurements can be easily accessible from either the vehicle's own wheel encoder or IMU. Covariance matrix $\boldsymbol{R}_k^{RSU}$ considers the uncertainty associated with the estimated position $\boldsymbol{x}_k^{RSU}$, which may be affected by the sequence length and noises in the range information. It is reported that the error of $\boldsymbol{x}_k^{RSU}$ is less than one meter. In our study, we conduct sensitivity analysis in Section VI to test whether RSU-assisted location accuracy will play a role in detecting and correcting GPS spoofing attacks.

Lastly, the latency needs to be considered when implementing the AES scheme to set up the secure channel between an RSU and vehicles. Here latency stems from three sources: the communication latency, the latency due to encrypting and decrypting the transmitted data, and the computational time to derive the vehicle's global position. One main contribution to the communication latency is the V2X technology involved, such as the Dedicated Short Range Communication (DSRC) and the emerging 5G-based Cellular-V2X (C-V2X) system. Previous studies have reported that the DSRC communication latency ranges from 10ms to 100ms [51], [52] and the C-V2X communication latency would not exceed 60ms even when there are 150 vehicles in the same communication channel [51], [53]. In our implementation, the run times for encrypting/ decrypting the transmitted data and deriving vehicle's global position are negligible (0.60ms and 0.13ms, respectively), when evaluated from an average of 1000 runs on a personal computer (with a 3.60GHz AMD Ryzen 7 CPU). This suggests that the latency of the designed secure RSU data is dominated by the communication latency. In this paper, we use 100ms, the largest reported communication latency in the numerical experiments.

2) **RSU-based location prediction**

The relative vehicle position measured by RSU, $z_k^{RSU}$, would not always be available, depending on the availability of RSUs along the road. Due to budget limits in a real-world setting, RSUs may be spatially sparse in the road network and RSU data is only available when vehicles are within an RSU's service range. In this study, we assume the distance between two consecutive RSUs, denoted as $D_{RSU}$, is uniform, and the service range $d_{RSU}$ is fixed. In Section VI, we conduct sensitivity analyses on how the spacing of RSUs will impact the performance of the proposed methods.

If RSU data are unavailable, we utilize the last available RSU data and vehicle motion model to predict a vehicle's location, enabling us to continuously monitor GPS measurements and timely detect attacks. The prediction should not involve GPS measurements that may have been compromised at the time when attacks are detected. However, since the vehicle location may change dramatically following commands from the vehicle's actuator (e.g., throttle, brake and steer), predicting the vehicle location can be challenging.

We build an RSU-based prediction model leveraging RSU data and the vehicle motion model to address this challenge. Specifically, given the most recent vehicle (global) position information ($\boldsymbol{x}_k^{RSU}$; see **Error! Reference source not found.**)) enabled by the RSU at time $k$, we predict vehicle location at $k + \Delta k$. For this, we start a standalone vehicle motion model at $k$, initialize it with $\boldsymbol{x}_k^{RSU}$ and then iterate it using IMU data $\boldsymbol{u}_t$ ($t \in [k+1, k+\Delta k]$) as the input. Note that besides predicting vehicle locations, we also propagate the errors in IMU data to gain the prediction uncertainty that is represented by a covariance matrix $\boldsymbol{P}_t^{RSU}$. The iterations of $\boldsymbol{x}_t^{RSU}$ and $\boldsymbol{P}_t^{RSU}$ are expressed in (10). We will use this prediction model in Section V-B to detect GPS spoofing and in Section V-C to correct the vehicle location when GPS spoofing is detected.

$$
\begin{aligned}
\boldsymbol{x}_t^{RSU} &= f\left(\boldsymbol{x}_{t-1}^{RSU}, \boldsymbol{u}_k\right) \\
\boldsymbol{P}_t^{RSU} &= \boldsymbol{F}_{t-1}\boldsymbol{P}_{t-1}^{RSU}\boldsymbol{F}_{t-1}^{T} + \boldsymbol{L}_{t-1}\boldsymbol{Q}\boldsymbol{L}_{t-1}^{T} \\
&t \in [k+1, k+\Delta k]
\end{aligned} \tag{10}
$$

Here, $\boldsymbol{F}_{t-1} = \frac{\partial f_{t-1}}{\partial x_{t-1}}|_{x_{t-1}^{RSU}}$ and $\boldsymbol{L}_{t-1} = \frac{\partial f_{t-1}}{\partial w_{t-1}}|_{x_{t-1}^{RSU}}$ are the partial derivative matrices w.r.t. the state $\boldsymbol{x}$ and IMU noises $\boldsymbol{w}$.

### *B. iForest Model-base Attack Detection*

Given the RSU data, the spoofing detection is formulated as a real-time anomaly detection problem, containing two parts: 1) generating real-time features, and 2) building a machine learning model that determines whether a GPS measurement is anomalous or not given the features at $k$.

1) **Feature generation**

- *The classical feature NEES*

We start with the classical feature for GPS spoofing detection, called NEES (Section II-B). It is computed as the normalized deviation of the received (possibly spoofed) GPS $\tilde{\boldsymbol{z}}_k^{GPS}$ from the predicted location $\hat{\boldsymbol{x}}_k$, denoted as $\boldsymbol{r}_k^{GPS}$, as below.

$$
\begin{aligned}
\boldsymbol{r}_k^{GPS} &= \tilde{\boldsymbol{z}}_k^{GPS} - \boldsymbol{H}\hat{\boldsymbol{x}}_k \\
\boldsymbol{S}_k^{GPS} &= \boldsymbol{H}\hat{\boldsymbol{P}}_k\boldsymbol{H}^{T} + \boldsymbol{R}^{GPS} \\
NEES_k^{GPS} &= \left(\boldsymbol{r}_k^{GPS}\right)^{T}\left(\boldsymbol{S}_k^{GPS}\right)^{-1}\boldsymbol{r}_k^{GPS}
\end{aligned} \tag{11}
$$

Note that $\boldsymbol{H}$ and $\widehat{\boldsymbol{P}}_k$ are defined in Section III-C, and $\boldsymbol{S}_k^{GPS}$ is a covariance matrix reflecting the uncertainty of $\boldsymbol{r}_k^{GPS}$.

It has been proven that if the noises in measurements follow a normal distribution, NEES follows a $\chi^2$ distribution [34]. Therefore, in previous studies, the $\chi^2$-test-based detection using $NEES_k^{GPS}$ is often applied to detect GPS spoofing. However, NEES could be impacted by noisy GPS measurements, making it hard to differentiate attacks from noises [18]. Furthermore, the $\chi^2$-test-based detection could be ineffective for stealthy attacks [1]. This is because attackers could inject a sequence of false information into the authentic GPS measurements; each piece of false information alone may not lead to a large enough NEES to trigger the alarm, but these errors together could successfully deviate the vehicle. If this happens, the $\chi^2$-test-based detector itself may also be compromised, making it less likely to detect spoofing attacks.

- *Features generated from RSU data*

We can create new features based on the measurements from RSUs, without involving GPS measurements, to address the issues associated with NEES. A straightforward way to create new features is to compute the difference between RSU and GPS measurements. However, as noted earlier, measurements from RSUs and GPS may not be at the same frequency, with the former not always being available. As a result, the two would not be directly comparable.

We address this issue by utilizing the RSU-based location prediction (see Section V-A). The predicted location is generated whenever a GPS measurement is received and needs to be validated. Then, new features are created by comparing the GPS measurement with RSU-based prediction in (10). Since the prediction in (8) does not involve GPS measurements, these features are 'protected' as they are immune to GPS spoofing attacks. Specifically, using the RSU-based location prediction $\boldsymbol{x}_k^{RSU}$ and the associated covariance matrix $\boldsymbol{P}_k^{RSU}$ (see Section V-A), we first compute the residual between the GPS measurement and the prediction $\boldsymbol{r}_k^{GPS}$ as well as the uncertainty of the residual $\boldsymbol{S}_k^{RSU}$, following (12). Then we generate two new (scalar) features $r_k^{RSU}$ and $S_k^{RSU}$, as shown in (13).

$$\begin{aligned} \boldsymbol{r}_k^{RSU} &= \tilde{\boldsymbol{z}}_k^{GPS} - \boldsymbol{H}\boldsymbol{x}_k^{RSU} \\ \boldsymbol{S}_k^{RSU} &= \boldsymbol{H}\boldsymbol{P}_k^{RSU}\boldsymbol{H}^T + \boldsymbol{R}^{RSU} \end{aligned} \tag{12}$$

$$\begin{aligned} r_k^{RSU} &= \left\|\boldsymbol{r}_k^{RSU}\right\| \\ S_k^{RSU} &= \left|\boldsymbol{S}_k^{RSU}\right| \end{aligned} \tag{13}$$

Here, $\|\bullet\|$ and $|\bullet|$ compute the L2 norm of a vector and the determinant of a matrix, respectively.

2) **Building an Isolation Forest as the detector**

The attack detection is treated as a real-time anomaly detection problem, for which we apply an unsupervised machine learning model to learn anomalies from the data. Specifically, we detect GPS spoofing by building an Isolation Forest (iForest) that takes all the above features $\boldsymbol{A}_k = (NEES_k^{GPS}, r_k^{RSU}, S_k^{RSU})$ at time k as the input. Note that though $NEES_k^{GPS}$ may be corrupted due to GPS spoofing and thus not a reliable feature alone, valuable information can be generated by comparing it with the other features, providing additional dimensions of inconsistency (anomaly) check.

iForest produces binary outputs: $\delta_k = 1$indicates being under attack and $\delta_k = -1$ indicates otherwise. Compared with other unsupervised learning methods, iForest has multiple advantages [54]. First, it has shown superior performance in detecting anomalies in extensive empirical studies. Second, iForest is easy to train in terms of selecting hyperparameters and can scale up to massive applications due to its linear time complexity and low memory consumption, making it suitable to run on vehicles with constrained resources.

The intuition behind iForest is that anomalous (or malicious) samples are easier to separate (i.e., isolate) from others compared with benign samples. In order to isolate a sample, the algorithm recursively generates partitions on all the samples by randomly setting a split (e.g., a threshold with a random feature) until all samples are separated. The recursive partitioning process is represented by growing a tree structure named Isolation Tree (iTree), with the leaves (or terminating nodes) being separated samples and intermediate nodes being attribute splits. Then, the length of the path to reach a sample starting from the root of an iTree approximates the number of partitions required to isolate the sample; a short length suggests a sample suspicious to be anomalous (as it is easier to separate). By constructing a large number of (random) iTrees based on the training dataset, we build an iForest. Using this iForest, we can identify samples that tend to have shorter path lengths in iTrees than others as anomalous. Anomaly detection with iForest consists of two stages: 1) a training dataset is used to build a forest of iTrees (i.e., iForest), and 2) each testing sample is passed through these iTrees, and an average anomaly score is assigned to the sample, which is further classified as a binary value. Readers are referred to [54] for more details.

An unsupervised learning method, the iForest can be trained without labeling the data; thus, the training data can be easily prepared. In this study, we generate training samples by running vehicles and collecting the features at each time step. It is worth noting that iForest works in scenarios where the training dataset does not contain any anomalies.

Therefore, we could prepare training samples using historical data, which may or may not be attacked. In this study, the training data is collected by running vehicles without GPS spoofing. The trained iForest can then be applied to detect GPS spoofing attacks in real-time. As expressed by (14), to check whether the GPS measurement at time $k$ is spoofed, we compute a set of real-time features $\boldsymbol{A}_k$ and input them to the trained iForest. An attack is detected if $\delta_k = 1$.

$$\delta_k = iForest(\boldsymbol{A}_k)\ ,\ \ \delta_k \in \{-1, 1\}\ . \tag{14}$$

In applications where GPS noise is large, we improve the robustness of the iForest-based detector by accounting for the temporal pattern of the features. Specifically, we apply a sliding window to use not only the features at time $k$ but also the ones at the previous time steps. In our experiment study where GPS noises are assumed large, features at the previous two steps (i.e., $\boldsymbol{A}_{k-2}, \boldsymbol{A}_{k-1}$) are incorporated to detect attacks at time $k$, as it is not common to observe three outliers consecutively (15). One may adopt a wider sliding window at the cost of a higher false-negative rate.

$$\delta_k = iForest(\boldsymbol{A}_{k-2}, \boldsymbol{A}_{k-1}, \boldsymbol{A}_k)\ ,\ \ \delta_k \in \{-1, 1\}\ . \tag{15}$$

Note that using the new features calculated from RSU data, similar machine learning methods, such as OCSVM (see Section II), can also be used to develop the detector, with their specific challenges addressed properly (e.g., choosing proper kernel functions and associated parameters for OCSVM [55]).

### *C. Infrastructure-enabled Correction*

Measurements from RSUs can also be used to correct vehicle positions, which is triggered either (a) when RSU data is received, or (b) when the detector detects GPS spoofing; see Fig. 2. In (b), the RSU-based location predictions will be used for correction if a vehicle is outside of the service range of RSUs. We introduce each case in detail in the following.

#### 1) **When RSU data is received**

When a vehicle enters the service range of an RSU, the vehicle periodically obtains measurements from the RSU, which can be used to correct the location estimation. The correction is done by directly initializing the state of EKF ($\widehat{\boldsymbol{x}}_k, \widehat{\boldsymbol{P}}_k$) following (16).

$$\begin{aligned} \widehat{\boldsymbol{x}}_k &= \boldsymbol{x}_k^{RSU} \\ \widehat{\boldsymbol{P}}_k &= \boldsymbol{P}_k^{RSU} \end{aligned} \tag{16}$$

Here, $\left(\boldsymbol{x}_k^{RSU}, \boldsymbol{P}_k^{RSU}\right)$ is the secure location estimation from RSU data in **Error! Reference source not found.**). An alternative way to correct vehicle position using RSU data is to follow the EKF's update step as introduced in Section III-C. However, this may not be reliable in stealthy attacks, which may bypass the attack detector and gradually corrupt the EKF [1]. The proposed method can effectively remove an attack's negative effects via direct initialization.

#### 2) **When GPS spoofing is detected**

When the detector detects an attack, besides isolating the GPS sensor, it corrects the EKF estimator as well. If RSU data is available, (16) is followed to correct the EKF location estimator; if not, the predicted location from the RSU-based prediction model is used. Specifically, when GPS spoofing is detected starting at $k + \Delta k$ but RSU data is not available, the predicted position $\boldsymbol{x}_{k+\Delta k}^{RSU}$ and its covariance matrix $\boldsymbol{P}_{k+\Delta k}^{RSU}$ in (10) are used by directly initializing the EKF state.

Since the RSU-based prediction model does not involve GPS measurements that may have been spoofed, the predicted location is able to correct errors resulting from a delayed detection, where the EKF estimator may have been compromised already. We show in Section VI-B that this brings benefits in defending stealthy attacks that are often detected with a delay. This is distinctively different from existing spoofing defense methods without RSU data: without removing the negative effect of the spoofed GPS measurements, they tend to generate deviated location estimation even if the vehicle successfully detects and isolates falsified GPS measurements.

## VI. EXPERIMENTAL STUDY

### *A. Experiment Settings*

#### 1) **General settings**

We test the proposed IED framework using both simulation data and real-world data. Simulation data is from the Downtown Seattle simulation model (Fig. 3a) built in Simulation Urban Mobility (SUMO). Fifty-three passenger vehicles are randomly selected for testing. Their trajectories allow us to capture diverse driving scenarios, including highways and local streets, where road geometries and vehicle dynamics vary considerably. The real-world GPS data contains trajectories from 15 vehicles, including both delivery trucks and passenger cars. Passenger car trajectories were collected from two field experiments conducted in Albany, NY, which were originally for measuring traffic performance [56]. Truck trajectories were provided by several anonymous logistic companies. Each vehicle trajectory comprises a sequence of time, location and speed reports, collected every 1 s.

Taking a trajectory as the input, the MATLAB Navigation Toolbox (MNT) is used to simulate necessary sensor measurements along the trajectory, including local (e.g., IMU), global (e.g., GPS data) and range measurements (e.g., RSU data). GPS data are manipulated following the attack models (Section IV) to simulate GPS spoofing attacks. The parameters of the IMU and GPS sensors (e.g., accuracy levels and resolutions) are set as MNT's default values, which reflect real-world sensor properties to a large extent. See details of MNT's sensor models in MATLAB documentation [57]. IMU and GPS measurements are sampled at 10Hz and 1Hz, respectively. RSUs are located along the road at an equal distance, and the service range of an RSU is represented by a circle with a radius of 500 meters centering at the RSU. Under the service range of an RSU, radio signal-to-noise ratio (SNR) in dB is simulated using the ground-truth range (i.e., the distance between the vehicle and the RSU) and following the measurement model $SNR = 10\log_{10}(|\boldsymbol{z}^{RSU}|^2/(\sigma^{RSU})^2)$ as in [50] (essentially the reverse of the ranging method). Here, $|\boldsymbol{z}^{RSU}|$ is the Euclidean distance between the vehicle and RSU and $\sigma^{RSU}$ represents the uncertainty (see Section V-A), which will be investigated further in our sensitivity analysis. The encryption, decryption and transmission process for data security is simulated via an AES scheme assuming a 100ms latency as noted earlier.

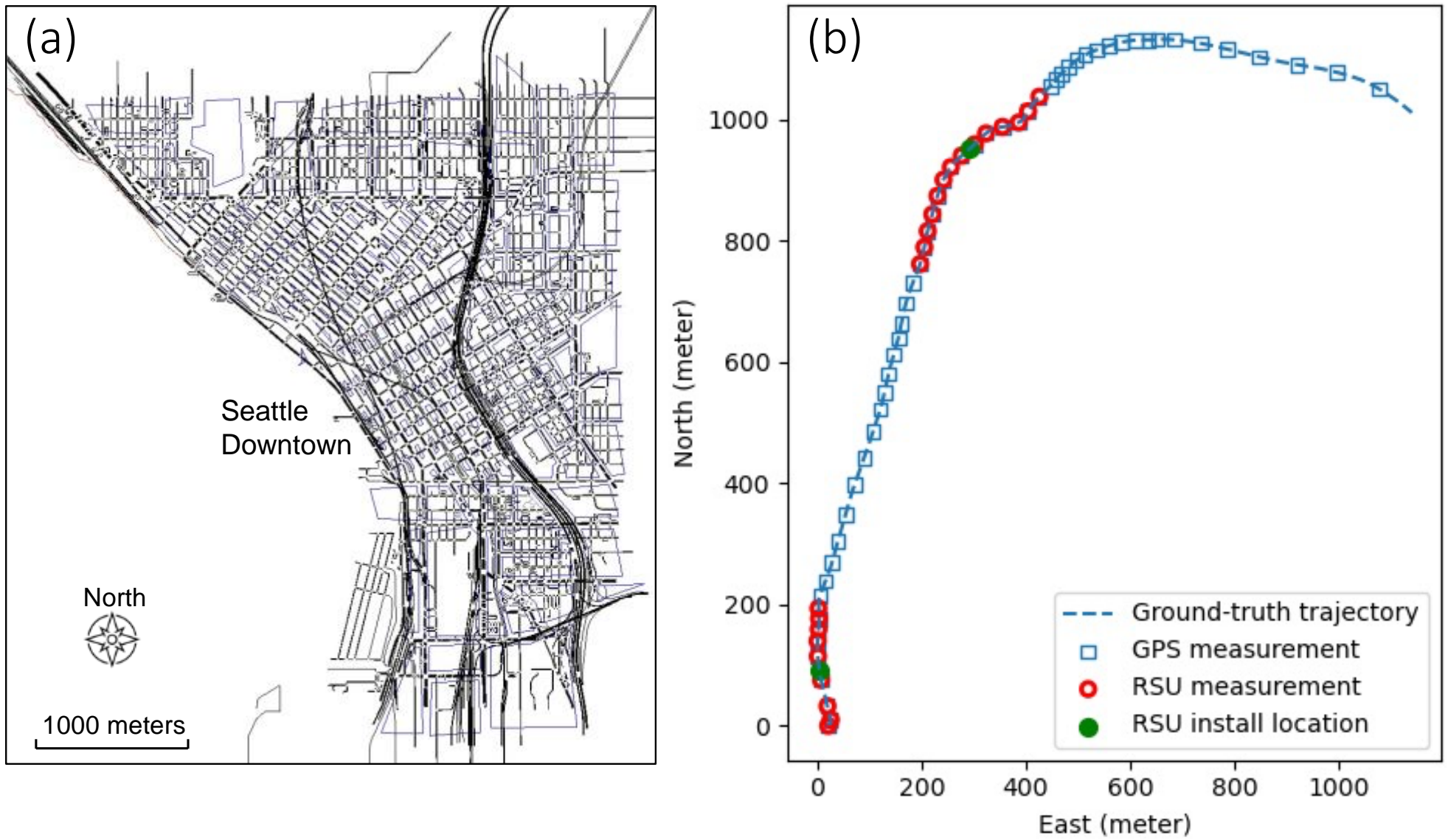

**Fig. 3.** (a) ROAD NETWORK OF DOWNTOWN SEATTLE IN SUMO; (b) SENSOR MEASUREMENTS ALONG A TRAJECTORY.

Fig. 3b illustrates an example of a ground-truth trajectory together with IMU, GPS and RSU measurements to help understand the sensor data. The measurements are visualized at where they are received/computed. It can be observed that GPS measurements are periodically received along the trajectory while RSU measurements are not spatially continuous but clustered around where RSUs are installed.

To simulate GPS spoofing, we randomly select the start time and duration of an attack (a uniform distribution ranging from 5 to 35 seconds). The attack modifies the true GPS data and passes the modified data to the vehicle for location estimation. We simulate the two types of attacks in Section IV. For constant bias attacks, GPS measurements are modified to deviate them by four meters, which is roughly the lane width. For stealthy attacks, *m=1.0* and *n=1.07*, respectively. We choose the two values so that the maximum deviation is comparable with the one in the constant bias attack (e.g., four meters) for an average attack duration of 20 seconds. These values also approximate the ones used in the original study [1].

**2) Overview of the experiments and key metrics**

We compare the iForest-based IED method with benchmark methods (see Section II) that include the $\chi^2$-test-based detector, the CUSUM detector, the OCSVM detector implemented following [13] without using RSU data, and the OCSVM detector that uses the RSU data (hereafter referred to as the "OCSVM-based IED method" as it uses the same set of features as the iForest method). For a detector that requires tuning parameters, we search around the parameters suggested in the original work and take the ones that yield the best performance in our experiments. The performance is averaged over all tested trajectories with random attack times (i.e., random attack starting time and attack duration).

TABLE I
PERFORMANCE OF THE PROPOSED AND BENCHMARK METHODS ON SIMULATED TRAJECTORIES

| | | $\chi^2$-test-based | CUSUM | Conventional OCSVM (No RSU data) | OCSVM-based IED | iForest-based IED |
|---|---|---|---|---|---|---|
| Constant bias attack | *F1 Score* | 0.56 | 0.69 | 0.70 | 0.83 | 0.86 |
| | *Precision* | 0.52 | 0.60 | 0.55 | 0.72 | 0.77 |
| | *Recall* | 0.69 | 0.86 | 0.95 | 0.98 | 0.99 |
| | *Detection lag* | 0 | 0 | 0 | 0 | 0 |
| | *RMSE* | 5.74 | 4.53 | 5.02 | 0.36 | 0.43 |
| Stealthy attack | *F1 score* | 0.47 | 0.21 | 0.48 | 0.72 | 0.78 |
| | *Precision* | 0.53 | 0.30 | 0.45 | 0.62 | 0.76 |
| | *Recall* | 0.49 | 0.22 | 0.57 | 0.86 | 0.84 |
| | *Detection lag* | 3 | 14 | 4 | 2 | 2 |
| | *RMSE* | 5.69 | 4.58 | 5.01 | 0.36 | 0.42 |

The similarities and differences of results from simulated and real-world trajectories are summarized. Lastly, we present results from sensitivity analysis of the iForest-based IED method under three influential factors, i.e., RSU spacing $D^{RSU}$, hyperparameter of iForest $\alpha$, and the error in RSU-assisted localization $\sigma^{RSU}$.

Several common metrics are adopted to evaluate the performance of the methods, including the *F1* score, *precision*, *recall*, *detection lag*, and *Rooted Mean Square Error* (*RMSE*) of location estimations. The first three evaluate detection accuracy, ranging from 0 to 1. *Precision* calculates the ratio of true positives (TP) over all the identified positives, and *recall,* also termed as the *correct detection probability*, is the ratio of true positives (TP) to all ground-truth positives. A higher precision and recall mean a lower false-positive (FP) rate and a higher TP rate, respectively. A higher *F1* means better detection performance in terms of balancing FP and TP.

$$F1 = \frac{2*precision*recall}{precision+recall} \tag{17}$$

Starting from the beginning of a spoofing, *detection lag* counts the number of GPS measurements (at 1Hz) missed by the detector before the attack is detected. If GPS measurements are spoofed but not detected timely, the victim vehicle assimilates them for location estimation, leading to a deviated trajectory. *RMSE* measures the location estimation error along a trip by computing the distance between the estimated locations $\hat{\boldsymbol{x}}_k$ and true locations $\boldsymbol{x}_k$:

$$RMSE = \sqrt{\frac{1}{K}\sum_{k=0}^{K}\|\hat{\boldsymbol{x}}_k - \boldsymbol{x}_k\|_2} \tag{18}$$

Here, $\|\hat{\boldsymbol{x}}_k - \boldsymbol{x}_k\|_2$ is the distance between the true location and estimated location at time *k*. *K* is the duration of the trajectory.

### *B. Testing Results using Simulated Trajectories*

We first evaluate the proposed method using the simulated trajectories; the results are reported in Table I.

Under the constant bias attacks, it can be found that all the methods can detect the start of attacks with no lag. Yet, given its strength in balancing FP and false negatives (FN) errors, the IED methods give F1 scores of 0.86 and 0.83, respectively, which are much better than the other three non-IED methods. The precision and recall of the IED methods indicate that they could nearly identify all the spoofed GPS measurements while generating some FPs possibly due to noises in GPS sensors. As shown later in the sensitivity analysis, we can reduce FNs while curbing FPs by tuning the hyperparameter. On the other hand, the low precisions by the three non-IED methods suggest that they produce many FPs. The conventional OCSVM has high recalls with low precisions, as it tends to produce FPs. The performance of OCSVM-based IED method, by utilizing RSU data, can be boosted significantly, which is similar to the performance of the iForest-based IED method. OCSVM has a lower F1 score due to its sensitivity to outliers as discussed in Section II-B. The results, especially the similar performances between iForest-based and OCSVM-based IED methods, suggest it is the new features computed from infrastructure data, not the specific learning methods, that lead to the improved performances of the IED methods. With effective detection and correction, the IED methods can dramatically reduce location errors compared to non-IED methods.

Under stealthy attacks, the IED methods give F1 scores of 0.78 and 0.72 respectively, again much better than the

TABLE II

PERFORMANCES OF THE PROPOSED AND BENCHMARK METHODS ON REAL-WORLD TRAJECTORIES

| | | $\chi^2$-test-based | CUSUM | Conventional OCSVM (No RSU data) | OCSVM-based IED | iForest-based IED |
|---|---|---|---|---|---|---|
| Constant bias attack | *F1 Score* | 0.54 | 0.66 | 0.64 | 0.80 | 0.93 |
| | *Precision* | 0.64 | 0.61 | 0.48 | 0.67 | 0.88 |
| | *Recall* | 0.52 | 0.75 | 0.95 | 1.00 | 1.00 |
| | *Detection lag* | 0 | 0 | 0 | 0 | 0 |
| | *RMSE* | 2.52 | 1.90 | 2.04 | 0.12 | 0.17 |
| Stealthy attack | *F1 score* | 0.44 | 0.22 | 0.45 | 0.69 | 0.73 |
| | *Precision* | 0.64 | 0.54 | 0.38 | 0.67 | 0.83 |
| | *Recall* | 0.38 | 0.14 | 0.58 | 0.76 | 0.67 |
| | *Detection lag* | 6 | 16 | 6 | 4 | 6 |
| | *RMSE* | 2.02 | 1.19 | 1.90 | 0.17 | 0.17 |

three non-IED methods. These findings suggest that the IED methods can effectively detect the attacks, despite the fact that the measurements from RSUs are not always available. Some interesting findings can be observed by comparing the performance under the two types of attacks. First, it is reasonable to observe that all the tested methods perform worse under stealthy attacks. Noteworthy is that though being downgraded, the IED's performance under stealthy attacks is still promising: the recall of 0.84 (or 0.86 for OCSVM-based IED) suggests that 84% (or 86%) of spoofed GPS measurements can be successfully detected. Second, unlike constant bias attacks, all the tested methods experience detection lags under stealthy attacks. Both IED methods miss two spoofed GPS measurements, as indicated by the detection lag in Table I. This is due to the attacks' stealthy design, where added perturbations are small at the early stage of attacks. Although missed by the detector, the two spoofed GPS measurements only bring small deviations to the location estimation, which are corrected once attacks are detected.

*C. Testing Results using Real-world Trajectories*

We further evaluate the proposed methods using real-world GPS trajectories, and the results are reported in Table II. It can be found that the IED methods still outperform the other three methods under both types of attacks. The F1 scores under stealthy attacks decrease, suggesting that all the methods are less effective compared with detecting constant bias attacks.

The F1 scores are close to those of the tests using simulated trajectories, suggesting the IED methods are also effective in dealing with real-world data. In detecting stealthy attacks, the IED methods have smaller recalls while larger precisions, compared with results from those on simulated data. The smaller recalls also lead to longer detection lags. In the following sensitivity analyses, we show that the trade-off between precision and recall can be adjusted according to the practical needs by varying hyperparameters of attack detectors.

*D. Sensitivity Analysis*

Multiple factors may impact the performance of the IED methods, such as the distance between two consecutive RSUs, the hyperparameters, and the accuracy of RSU-assisted localization. Here we conduct sensitivity analyses on how these factors influence the iForest-based IED method. Simulated data are used for the analysis unless noted otherwise.

1) **Distance between two consecutive RSUs**

Given its reliance on RSU data, the IED method is expected to be influenced by RSU's deployment strategy. Specifically, out of the RSU service range, the vehicle relies on RSU-based location prediction for attack detection and correction. Table III shows the performance of varying RSU distance $D_{RSU}$ under attacks. Note that we stop at 2000m as most of the trajectories are shorter than 2000m and a larger $D_{RSU}$ does not reduce the performance further. As expected, the performance (such as F1 score and RMSE) downgrades as $D_{RSU}$ increases. Yet, the iForest-based IED method still maintains an advantage over the benchmark methods as $D_{RSU}$ increases.

Fig. 4 and Fig. 5 show the sensitivity of the false alarm rate and recall (correct detection probability) with $D_{RSU}$ (and the other two factors as well). It can be observed that a larger $D^{RSU}$ leads to a larger false alarm rate for both attacks. Under constant bias attacks, the recall stays close to 1, and the detection lags are zero, suggesting that these attacks can be easily and timely identified regardless of $D^{RSU}$ Under stealthy attacks, increasing $D^{RSU}$ from 1000m to 1500m does not affect recall significantly, while a larger $D^{RSU}$ (at 2000m) drops it.

TABLE III

INFLUENCE OF RSU SPACING ON IFOREST-BASED IED METHOD

| $D_{RSU}$ = | | 1000m | 1500m | 2000m |
|---|---|---|---|---|
| Constant bias attack | *F1 Score* | 0.92 | 0.86 | 0.82 |
| | *Precision* | 0.86 | 0.77 | 0.72 |
| | *Recall* | 0.99 | 0.99 | 0.99 |
| | *Detection lag* | 0 | 0 | 0 |
| | *RMSE* | 0.10 | 0.43 | 0.54 |
| Stealthy attack | *F1 score* | 0.83 | 0.78 | 0.62 |
| | *Precision* | 0.83 | 0.76 | 0.64 |
| | *Recall* | 0.83 | 0.84 | 0.65 |
| | *Detection lag* | 3 | 2 | 6 |
| | *RMSE* | 0.10 | 0.42 | 0.60 |

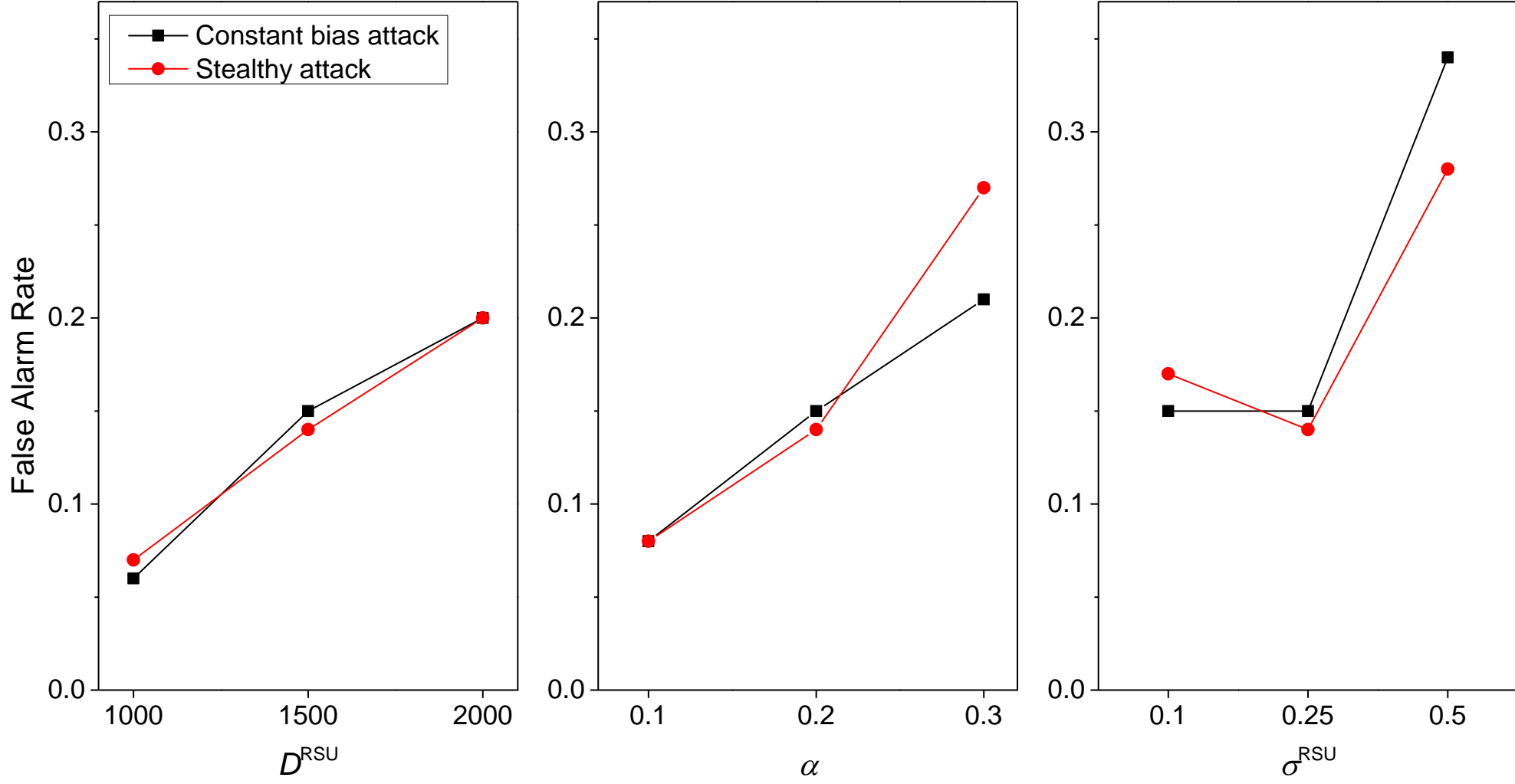

**Fig. 4.** INFLUENTIAL FACTORS AND FALSE ALARM RATE.

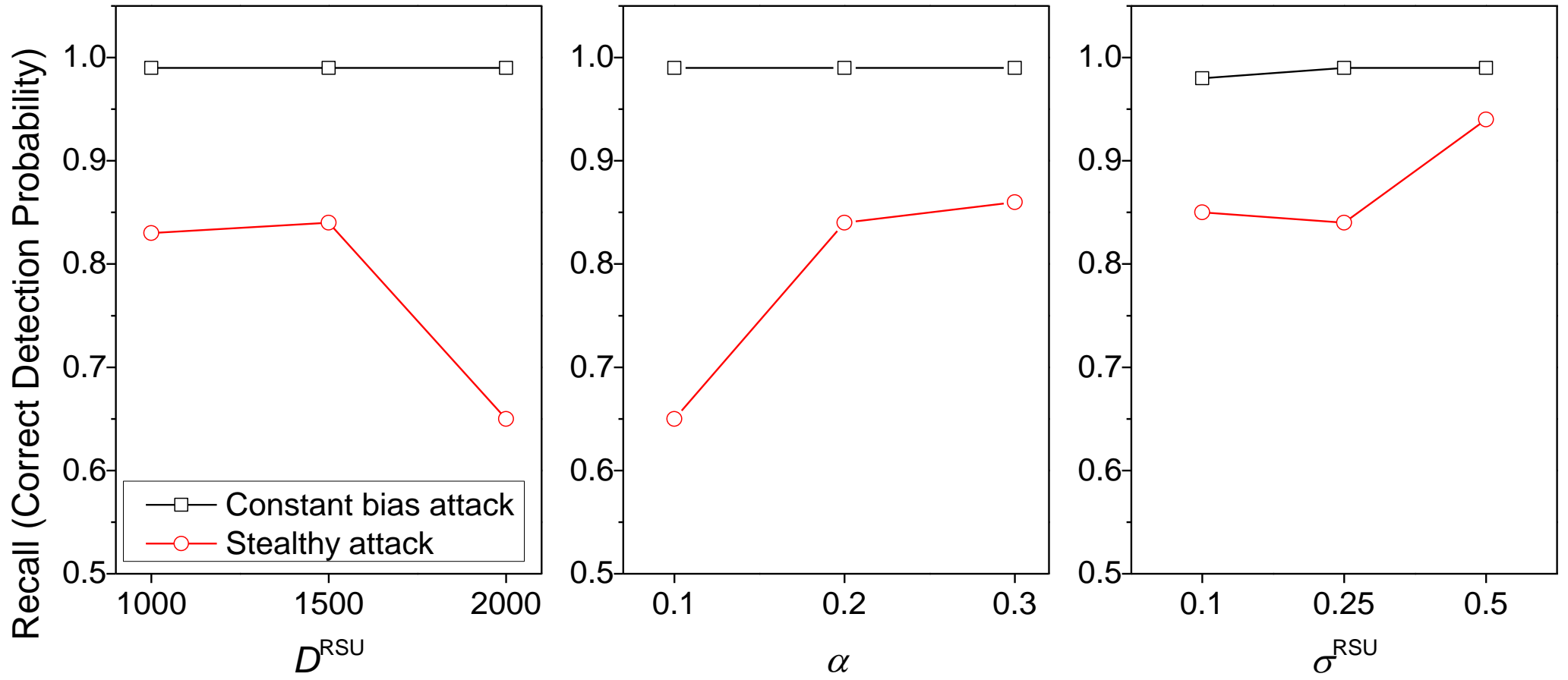

**Fig. 5.** EFFECTS OF INFLUENTIAL FACTORS ON RECALL.

2) **Hyperparameter of the attack detector**

In the iForest-based IED method, one key hyperparameter associated with iForest is contamination (denoted as $\alpha$)

which specifies the proportion of spoofed samples in the data set. Table IV summarizes the performance of the IED method with different $\alpha$, ranging from 0 to 0.5. 0 means no anomalies and 0.5 means that half of the data samples are anomalies. A range from 0.1 to 0.3 in Table IV captures a fairly large range of $\alpha$.

TABLE IV

IMPACTS OF IFOREST HYPERPARAMETER

| | $\boldsymbol{\alpha}$ = | 0.1 | 0.2 | 0.3 |
|---|---|---|---|---|
| Constant bias attack | *F1 Score* | 0.91 | 0.86 | 0.8 |
| | *Precision* | 0.86 | 0.77 | 0.69 |
| | *Recall* | 0.99 | 0.99 | 0.99 |
| | *Detection lag* | 0 | 0 | 0 |
| | *RMSE* | 0.25 | 0.43 | 0.31 |
| Stealthy attack | *F1 score* | 0.71 | 0.78 | 0.71 |
| | *Precision* | 0.8 | 0.76 | 0.62 |
| | *Recall* | 0.65 | 0.84 | 0.86 |
| | *Detection lag* | 6 | 3 | 2 |
| | *RMSE* | 0.35 | 0.42 | 0.45 |

Under the constant bias attacks, recalls remain unchanged (near 1), suggesting that the method can robustly detect spoofed GPS data under such attacks for a wide range of $\alpha$ (Table IV and Fig. 5). Meanwhile, a sensitive detector with a large $\alpha$ tends to reduce the detection lag. Yet, under both constant bias and stealthy attacks, a larger $\alpha$ leads to more FPs, as indicated by the increase in the false alarm rate (Fig. 4) and the decrease in precision (Table IV). On the other hand, a larger $\alpha$ brings benefits to detecting stealthy attacks, since i) more spoofed GPS measurements can be detected (as indicated by the larger recall), and ii) the detection lag is shorter.

In summary, a proper $\alpha$ can help balance precisions and recalls. The proper $\alpha$ depends on the types of attacks: a small $\alpha$ is good for detecting constant bias attacks but encourages FNs in stealthy attacks, reducing recalls. Given the high threat of stealthy attacks, it would be beneficial to set a relatively large $\alpha$ to effectively detect such attacks. In our experiments, a balance between precisions and recalls under the stealthy attack can be reached around $\alpha$=0.2.

3) **Accuracy of RSU-assisted localization**

Vehicle localization assisted by the RSU can be more accurate (in centimeters) than GPS measurements (in meters). In practice, the accuracy of RSU-assisted localization could depend on factors such as the ranging method applied, how RSUs are configured, and the real-time driving environments. Here, we check how the accuracy of RSU-assisted localization may impact the performance of the proposed method.

TABLE V

IMPACTS OF THE ACCURACY OF RSU-BASED LOCALIZATION

| | $\boldsymbol{\sigma^{RSU}}$ = | 0.1 | 0.25 | 0.5 |
|---|---|---|---|---|
| Constant bias attack | *F1 Score* | 0.86 | 0.86 | 0.71 |
| | *Precision* | 0.75 | 0.77 | 0.55 |
| | *Recall* | 0.98 | 0.99 | 0.99 |
| | *Detection lag* | 0 | 0 | 0 |
| | *RMSE* | 0.41 | 0.43 | 0.81 |
| Stealthy attack | *F1 score* | 0.77 | 0.78 | 0.72 |
| | *Precision* | 0.71 | 0.76 | 0.59 |
| | *Recall* | 0.85 | 0.84 | 0.94 |
| | *Detection lag* | 3 | 3 | 0 |
| | *RMSE* | 0.42 | 0.42 | 0.54 |

Table V shows three accuracy levels of RSU-based localization obtained by tuning the uncertainty parameter $\sigma^{RSU}$ (Section V-A). $\sigma^{RSU}$=0.5 means that about 95% of location errors are within one meter, which is often considered as the worst scenario for RSU-assisted localization [48]. It can be observed that compared with the baseline ($\sigma^{RSU}$=0.25), the higher accuracy in RSU-assisted localization ($\sigma^{RSU}$=0.1) has nearly no effect on detecting constant bias attacks but does improve the performance of detecting stealthy attacks that add tiny deviations at the beginning of an attack. A lower location accuracy ($\sigma^{RSU}$=0.5) reduces the performance in both types of attacks, with a higher false alarm rate (Fig. 4) and leading to larger location estimation errors (in RSME).

## VII. Conclusion and Discussions

In this paper, we proposed an infrastructure-enabled defense (IED) framework that utilizes secure RSU data for detecting GPS spoofing and correcting location errors from the spoofing. Timely detection is achieved by designing and training an iForest model using real-time features computed from both RSU data and (possibly spoofed) GPS data. Once spoofing is detected, GPS data is isolated and the compromised vehicle locations are corrected using RSU data. Experimental results using both simulation and real-world GPS data demonstrated that the IED framework enhances timely detection and correction even when RSU data is not spatially continuous. We showed that the IED framework is effective in defending against state-of-the-art stealthy GPS spoofing models. Furthermore, sensitivity analyses produced insights into how RSU deployment, hyperparameters, and the accuracy of RSU-assisted localization impact the IED's performance.

The IED framework for GPS spoofing distinguishes itself from non-IED methods in three major aspects. First, it relaxes the requirement of vehicular sensors, making detectors more robust when dealing with spoofing attacks. Second, enabled by the secure RSU data, a relatively simple detector based on an unsupervised learning algorithm (e.g., iForest or OCSVM) can effectively detect GPS spoofing attacks. The advantage stems from the fact that the features computed from secure RSU data for attack detection are "protected", relieving the challenges of developing attack-resilient algorithms. This advantage could be exploited to defend against false data injection attacks in general since the GPS spoofing setting adopted in this paper is general and can represent other false data injection attacks [1], [9]. That is, if the observation deviates too much from "the expected value" that is computed using secure infrastructure data, the observation is likely under attack.

Several limitations of the proposed IED framework call for future research. First, the detection and correction methods may be enhanced by more advanced learning approaches (such as deep learning) to further improve their performances. Second, more research efforts are needed to design optimal strategies for deploying the RSUs. In this study, we assumed RSUs are deployed evenly on the roadside and conducted sensitivity analyses to understand the impact of the distance between two consecutive RSUs on the IED's performance. For future research, the optimal RSU deployment problem may be studied to produce RSU deployment strategies that systematically consider the deployment cost, traffic environments, road geometry, and the performance of the spoofing defense method. Third, future investigations are needed to test the IED's performance in real-world driving scenarios where GPS spoofing attacks, infrastructure (RSU and implementation of the ranging method in Section V-A), and the IED framework are implemented and tested. Fourth, the IED framework may be enhanced by incorporating additional (and easily obtained) data sources for more robust location estimation and/or attack detection. This is particularly so for scenarios where the distance between RSUs is large. For instance, the geometric outlines of roads may be used as constraints to improve location estimation/prediction, which may further improve detection accuracy. Last but not least, as infrastructure is becoming more important in transportation, the idea of the proposed IED framework may be applied to other applications. This may include vehicular computer vision systems that are vulnerable to data attacks, e.g., adding adversarial images to onboard cameras [58], or spoofing attacks on LiDAR data [4]. The proposed IED framework may be applied to these applications by i) designing specific secure infrastructure data including what data to collect and how to secure data transmission, ii) computing new features from the infrastructure data to help develop effective attack detection methods, and iii) correcting possibly corrupted data by using infrastructure data. The authors will pursue these research directions, and results may be reported in subsequent papers.